\documentclass[prl,superscriptaddress,floatfix,twocolumn,aps,10pt,showpacs]{revtex4-1}

\usepackage{graphicx}
\usepackage{verbatim}
\usepackage{comment}
\usepackage{amssymb}
\usepackage{stmaryrd}
\usepackage{color}
\usepackage{mathptmx}
\usepackage{amsmath}
\usepackage[normalem]{ulem}

\newcommand{\WFini}{\vert \phi_\text{ini}\rangle}

\newcommand{\Eiga}{\vert\psi_\alpha\rangle}
\newcommand{\CEiga}{\langle\psi_\alpha\vert}

\newcommand{\appropto}{\mathrel{\vcenter{
  \offinterlineskip\halign{\hfil$##$\cr
    \propto\cr\noalign{\kern2pt}\sim\cr\noalign{\kern-2pt}}}}}

\newcommand{\be}{\begin{eqnarray}}
\newcommand{\ee}{\end{eqnarray}}
\def\beq{\begin{equation}}
\def\eeq{\end{equation}}

\begin{document}

\title{Fluctuation-Dissipation Theorem in an Isolated System of Quantum Dipolar Bosons after a Quench}
\author{Ehsan Khatami}
\affiliation{Department of Physics, University of California, Santa Cruz, California 95064, USA}
\author{Guido Pupillo}
\affiliation{IPCMS (UMR 7504) and ISIS (UMR 7006), Universit\'e de Strasbourg and CNRS, Strasbourg, France}
\author{Mark Srednicki}
\affiliation{Department of Physics, University of California, Santa Barbara, California 93106, USA}
\author{Marcos Rigol}
\affiliation{Department of Physics, The Pennsylvania State University, University Park, Pennsylvania 16802, USA}


\begin{abstract}
We examine the validity of fluctuation-dissipation relations in isolated quantum systems 
taken out of equilibrium by a sudden quench. We focus on the dynamics of trapped 
hard-core bosons in one-dimensional lattices with dipolar interactions whose strength 
is changed during the quench. We find indications that fluctuation-dissipation relations 
hold if the system is nonintegrable after the quench, as well as if it 
is integrable after the quench if the initial state is an equilibrium state of a 
nonintegrable Hamiltonian. On the other hand, we find indications that they fail if the system 
is integrable both before and after quenching.
\end{abstract}

\maketitle

The fluctuation-dissipation theorem (FDT)~\cite{nyquist_28,onsager_31a,callen_welton_51} 
is a fundamental relation in statistical mechanics which states that typical 
deviations from the equilibrium state caused by an external perturbation (within the linear response regime) 
dissipate in time in the same way as random fluctuations. The theorem applies to both classical and quantum 
systems as long as they are in thermal equilibrium. Fluctuation-dissipation relations are not, 
in general, satisfied for out-of-equilibrium systems. In particular, if a system is isolated, it is not clear 
whether once taken far from equilibrium fluctuation-dissipation relations apply at any later time.
Studies of integrable models such as a Luttinger liquid \cite{a_mitra_11} and the
transverse field Ising chain \cite{l_foini_11} have shown that the use of
fluctuation-dissipation relations to define temperature leads to values of the
temperature that depend on the momentum mode and/or the frequency being
considered. More recently, Essler {\em et al.}~\cite{essler_12} have shown 
that for a subsystem of an isolated infinite system, the basic form of the FDT holds,
and that the same ensemble that describes the static properties also describes the dynamics.

The question of the applicability of the FDT to isolated quantum systems is
particularly relevant to experiments with cold atomic gases
\cite{bloch_dalibard_review_08,cazalilla_citro_review_11},
whose dynamics is considered to be, to a good approximation, unitary
\cite{trotzky_chen_12}. In that context, the description of observables after
relaxation (whenever relaxation to a time-independent
value occurs) has been intensively explored in the recent literature
\cite{cazalilla_rigol_10}. This is because, for isolated quantum 
systems out of equilibrium, it is not apparent that thermalization can take place. 
For example, if the system is prepared in an initial pure state $\WFini$ that
is not an eigenstate of the Hamiltonian $\hat{H}$ ($\hat{H}\Eiga=E_\alpha\Eiga$)
(as in  Ref.~\cite{trotzky_chen_12}), then the infinite-time average of the evolution 
of the observable $\hat{O}$ can be written as
$
\overline{\langle \hat{O}(t) \rangle}=\sum_{\alpha} |c_{\alpha}|^{2} O_{\alpha\alpha} \equiv O_\text{diag},
$
where $c_{\alpha}=\langle\psi_\alpha\WFini$, $O_{\alpha\alpha}=\CEiga\hat{O}\Eiga$, and we have assumed that the 
spectrum is nondegenerate. The outcome of the infinite-time average can be thought of as the prediction of 
a ``diagonal'' ensemble \cite{rigol_dunjko_08}. $O_\text{diag}$ depends on the 
initial state through the $c_{\alpha}$'s (there is an exponentially large number of them), while the thermal 
predictions depend only on the total energy $\langle\phi_\text{ini}|\hat{H}\WFini$; i.e., they need not agree.

The lack of thermalization of some observables, in the specific case of quasi-one-dimensional geometries 
close to an integrable point, was seen in experiments \cite{kinoshita_wenger_06} short-range and, at 
integrability, confirmed in computational \cite{rigol_dunjko_07} and analytical 
\cite{cazalilla_06} calculations. Away 
from integrability, computational studies have shown that few-body observables thermalize in general 
\cite{rigol_dunjko_08,rigol_09a,banuls_cirac_11,kollath_lauchli_07}, 
which can be understood in 
terms of the eigenstate thermalization hypothesis (ETH) \cite{deutsch_91,srednicki_94,rigol_dunjko_08}. We note 
that the nonintegrable systems studied computationally belong to two main classes of lattice models: 
(i) spin-polarized fermions, hard-core bosons, and spin models with short-range (nearest and next-nearest-neighbor) interactions \cite{manmana_wessel_07,rigol_dunjko_08,rigol_09a,
banuls_cirac_11} and (ii) the Bose-Hubbard model 
\cite{kollath_lauchli_07}. 

In this Letter, we go beyond these studies and report results that indicate that fluctuation-dissipation 
relations are also valid in generic isolated quantum systems after relaxation, while they fail at 
integrability. For that, we use exact diagonalization and study a third class of lattice models, 
hard-core bosons with dipolar interactions in one dimension~\cite{lahaye_menotti_07}.
The latter are of special interest as they describe experiments with quantum gases of magnetic atoms trapped 
in optical lattices \cite{billy_henn_12} as well as ground state polar molecules \cite{jin_ye_12}. 
Rydberg-excited alkali atoms \cite{saffman_walker_10} and laser-cooled ions \cite{mintert_wunderlich_01} 
may soon provide alternative realizations of correlated systems with dipolar 
interactions. The effect of having  power-law decaying interactions in the dynamics and 
description of isolated quantum systems after relaxation is an important and open question that
we address here.

The model Hamiltonian for those systems can be written as
\beq
\hat{H}= - J \sum_{j=1}^{L-1} \left( \hat{b}^{\dagger}_j \hat{b}^{}_{j+1} + \text{H.c.} \right) +
V\sum_{j<l} \frac{\hat{n}^{}_j \hat{n}^{}_{l}}{|j-l|^3} +g\sum_{j} x_j^2\, \hat{n}^{}_j
\label{eq:model}
\eeq
where $\hat{b}^{\dagger}_j$ ($\hat{b}^{}_j$) creates (annihilates) a hard-core boson 
($\hat{b}^{\dagger 2}_j= \hat{b}^2_j=0$) at site $j$, and $\hat{n}^{}_j=\hat{b}^{\dagger}_j\hat{b}^{}_j$ is the 
number operator. $J$ is the hopping amplitude, $V$ the strength of the dipolar interaction, $g$ the strength 
of the confining potential, $x_j$ the distance of site $j$ from the center of the trap, and $L$ the number 
of lattice sites (the total number of bosons $p$ is always chosen to be $p=L/3$). We set $J=1$ (unit of energy 
throughout this paper), $\hbar=k_{B}=1$, use open boundary conditions, and work in the 
subspace with even parity under reflection. 

We focus on testing a fluctuation-dissipation relation after a quench for experimentally 
relevant observables, namely, site and momentum occupations (results for the density-density 
structure factor are presented in Ref.~\cite{note3}). A scenario under which FDT holds 
in isolated systems out of equilibrium was 
put forward by one of us in Ref.~\cite{srednicki_99}. There, it was shown that after a quantum 
or thermal fluctuation (assumed to occur at time $t'$~\cite{note7}, which was treated as a
uniformly distributed
random variable), it is overwhelmingly likely that $O_{t'\pm t}= C_{\text{Fluc}}(t)O_{t'}$, 
where $O_{t}=\langle\hat{O}(t)\rangle$~\cite{note1}.  Formally, $C_{\text{Fluc}}(t)$ 
is related to the second moments of a probability distribution for $O_t$, 
$C_{\text{Fluc}}(t)=\overline{O_{t+t''}O_{t''}}/\overline{(O_{t''})^2}$, where infinite-time 
averages have been taken with respect to $t''$. Therefore, assuming that no 
degeneracies occur in the many-body spectrum or that they are unimportant, $C_{\text{Fluc}}(t)$ 
can be written as
\be
C_{\text{Fluc}}(t)\propto\sum_{\substack{\alpha\beta \\ \alpha\neq \beta}}|c_{\alpha}|^2|c_{\beta}|^2
|O_{\alpha\beta}|^2 e^{i(E_{\alpha}-E_{\beta})t},
\label{eq:fluc}
\ee
where the proportionality constant is such that $C_{\text{Fluc}}(0)=1$~\cite{note2}. 
The correlation function in Eq.~(\ref{eq:fluc}) explicitly depends on the initial 
state through $c_{\alpha}$.

Assuming that eigenstate thermalization occurs in the Hamiltonian of
interest, the matrix elements of $\hat{O}$ in the energy eigenstate basis
can be written as
\be
O_{\alpha\beta}=\Omega(E)\delta_{\alpha\beta}+e^{-S(E)/2}f(E,\omega)R_{\alpha\beta},
\label{eq:chaotic}
\ee
where $E\equiv \frac{1}{2}(E_{\alpha}+E_{\beta})$, $\omega\equiv E_{\alpha}-E_{\beta}$,
$S(E)$ is the thermodynamic entropy at energy $E$, $e^{S(E)}=E\sum_{\alpha}\delta(E-E_{\alpha})$,
$\Omega(E)$ and $f(E,\omega)$ are smooth functions of their arguments, and 
$R_{\alpha\beta}$ is a random variable (e.g., with zero mean and unit variance).
This is consistent with quantum chaos theory and is presumably valid for a wide range 
of circumstances~\cite{srednicki_96,srednicki_99}. From Eq.~\eqref{eq:chaotic}, it follows
straightforwardly that 
$
C_{\text{Fluc}}(t)\sim C_{\text{Appr}}(t),
\label{eq:fluc-appr}
$
where we have defined
\be
C_{\text{Appr}}(t)\propto\int_{-\infty}^{+\infty}d\omega|f(E,\omega)|^2e^{i\omega t},
\label{eq:appr}
\ee
and again, the proportionality constant is such that $C_{\text{Appr}}(0)=1$~\cite{note4}.
Therefore, we see that
$C_{\text{Fluc}}(t)$ does not depend on the details of the initial state, in the same way 
that observables in the diagonal ensemble do not depend on such details. 

We can then compare this result to how a typical deviation from thermal equilibrium 
(used to describe observables in the nonequilibrium system after relaxation) caused by an external 
perturbation ``dissipates'' in time. Assuming that the perturbation is small (linear response regime)
and that it is applied at time $t=0$, $C_{\text{Diss}}(t)$, defined via
$O_{t}= C_{\text{Diss}}(t)O_\text{Thermal}$,
can be calculated through Kubo's formula as~\cite{kubo57,srednicki_99}
\be
C_{\text{Diss}}(t)\propto\sum_{\substack{\alpha\beta \\ \alpha\neq \beta}}\frac{e^{-E_{\alpha}/T}
-e^{-E_{\beta}/T}}{E_{\beta}-E_{\alpha}}|O_{\alpha\beta}|^2 e^{i(E_{\alpha}-E_{\beta})t},
\label{eq:diss}
\ee
where again, we set $C_{\text{Diss}}(0)=1$.
Using Eq.~\eqref{eq:chaotic}, one finds that
\beq
C_{\text{Diss}}(t)\sim\int_{-\infty}^{+\infty}d\omega\frac{\sinh(\omega/2T)}{\omega}
|f(E,\omega)|^2e^{i\omega t}\sim C_{\text{Appr}}(t),
\label{eq:dissappr}
\eeq
where the last similarity is valid if the width of $f(E,\omega)$~\cite{note3}
is of the order of, or smaller than, the temperature. The results in Eqs.~\eqref{eq:appr} and 
\eqref{eq:dissappr} suggest that FDT holds in isolated quantum systems out of equilibrium under very 
general conditions.

In what follows, we study dipolar systems out of equilibrium and test whether their dynamics is 
consistent with the scenario above. This is a first step toward understanding the relevance of FDT
and of the specific scenario proposed in Ref.~\cite{srednicki_99}, to experiments with nonequilibrium 
ultracold quantum gases. The dynamics are studied after sudden quenches, for which the initial pure state $\WFini$ 
is selected to be an eigenstate of Eq. \eqref{eq:model} for $V=V_\text{ini}$ and $g=g_\text{ini}$ 
($\hat{H}_\text{ini}$), and the evolution is studied under $\hat{H}_\text{fin}$ ($V=V_\text{fin}$ 
and $g=g_\text{fin}$), i. e., $|\phi(t)\rangle=e^{-i\hat{H}_{\text{fin}}t}\WFini$. We 
consider the following three types of quenches: type (i) \{$V_\text{ini}=0$, $g_\text{ini}=\gamma$\}$\to$ 
\{$V_\text{fin}=0$, $g_\text{fin}=\gamma/10$\} (integrable to integrable),
type (ii) \{$V_\text{ini}=8$, $g_\text{ini}=\gamma$\}$\to$ \{$V_\text{fin}=0$, 
$g_\text{fin}=\gamma$\} (nonintegrable to integrable), and
type (iii) \{$V_\text{ini}=8$, $g_\text{ini}=\gamma$\}$\to$ \{$V_\text{fin}=2$, 
$g_\text{fin}=\gamma$\} (nonintegrable to nonintegrable).
We choose $\gamma$ such that $\gamma x_1^2=\gamma x_L^2=4$, 
which ensures a (nearly) vanishing density at the edges of the lattice in the ground state.
The initial state for different quenches, which need not be the ground state of $\hat{H}_\text{ini}$, 
is selected such that $E_{\textrm{tot}}=\langle\phi_\text{ini}|\hat{H}_\text{fin}\WFini$ corresponds to the energy 
of a canonical ensemble with temperature $T=5$, i.e., such that $E_{\textrm{tot}}=\text{Tr}\{e^{-\hat{H}_\text{fin}/T} 
\hat{H}_\text{fin}\}/\text{Tr}\{e^{-\hat{H}_\textrm{fin}/T}\}$. 

\begin{figure}[t]
\centerline {\includegraphics*[width=3.3in]{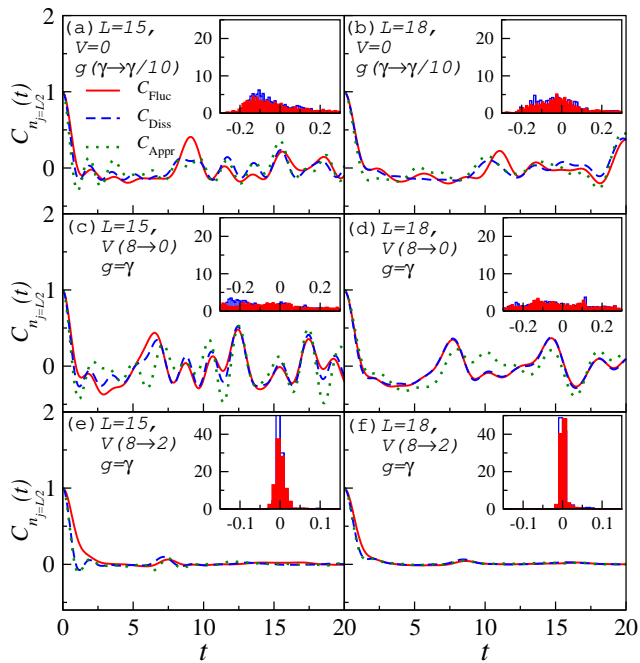}}
\caption{Correlation functions $C_{\text{Fluc}}(t),\ C_{\text{Diss}}(t),$ 
and $C_{\text{Appr}}(t)$ when the observable is $n_{j=L/2}$ vs time $t$. 
Results are shown for the three quenches (i)-(iii)  (from top to bottom, 
respectively) explained in the text, and for $L=15$ (left panels) and 18 (right panels). 
Results for $L=12$ are presented in Ref.~\cite{note3}. The insets show 
normalized histograms of $C_{\text{Fluc}}(t)$ (filled red bars) and 
$C_{\text{Diss}}(t)$ (empty blue bars) calculated for 2000 data points between 
$t=0$ and 100.}
\label{fig:T5-V8}
\end{figure}

In Fig.~\ref{fig:T5-V8}, we show results for $C_{\text{Fluc}}(t)$, $C_{\text{Diss}}(t)$, and 
$C_{\text{Appr}}(t)$ when the observable of interest is the occupation of the site in the 
center of the system $n_{j=L/2}$ (qualitatively similar results were obtained for other site 
occupations, for momenta occupations, and for the density-density structure factor~\cite{note3}). 
The results are obtained for the three different quench types mentioned above and are shown for 
$L=15$ and 18. For quench type (i), we find that none of the three 
correlation functions agree with each other and that the agreement does not improve
with increasing $L$ [see Figs.~\ref{fig:T5-V8}(a) and \ref{fig:T5-V8}(b)]. 
There are also large time fluctuations, characteristic of the 
integrable nature of the final Hamiltonian \cite{campos_zanardi_13}. We quantify these fluctuations 
by plotting the histograms of $C_{\text{Fluc}}(t)$ and $C_{\text{Diss}}(t)$ for an extended 
period of time in the insets. We find the histograms to be broad functions for quenches (i) and (ii)
[Figs.~\ref{fig:T5-V8}(a)-\ref{fig:T5-V8}(d)].

Remarkably, in quenches type (ii) [Figs.~\ref{fig:T5-V8}(c) and \ref{fig:T5-V8}(d)], which 
also have a final Hamiltonian that is integrable, $C_{\text{Fluc}}(t)$ and $C_{\text{Diss}}(t)$
are very similar to each other at each time and their differences decrease with
increasing $L$. This indicates that the FDT holds. At the same time, we find 
differences between fluctuation or dissipation correlations and $C_{\text{Appr}}(t)$, 
indicating that the agreement between $C_{\text{Fluc}}(t)$ and $C_{\text{Diss}}(t)$
does not imply that Eq.~\eqref{eq:chaotic} is valid.
These observations can be understood if the initial state provides an unbiased sampling
of the eigenstates of the final Hamiltonian. In that case, even though eigenstate 
thermalization does not occur, thermalization can take place \cite{rigol_srednicki_12}, 
and this results in the applicability of FDT. In quenches type (ii), such an unbiased 
sampling occurs because of the nonintegrability of the initial Hamiltonian, whose 
eigenstates are random superpositions of eigenstates of the final integrable Hamiltonian 
with close energies \cite{rigol_srednicki_12}.

For quenches type (iii) [Figs.~\ref{fig:T5-V8}(e) and \ref{fig:T5-V8}(f)], on the other hand,
we find that not only $C_{\text{Fluc}}(t)$ and $C_{\text{Diss}}(t)$ are very close to each 
other, but also $C_{\text{Appr}}(t)$ is very close to both of them, and that the 
differences between the three decrease with increasing $L$. Therefore, our results are 
consistent with the system exhibiting eigenstate thermalization \cite{note6}, which means that the assumptions 
made in Eq.~\eqref{eq:chaotic} are valid, and the applicability of the FDT follows.
Furthermore, for quenches type (iii), one can see that time fluctuations are strongly suppressed
when compared to those in quenches type (i) and (ii) [better seen in the insets of Fig.~\ref{fig:T5-V8}(e) and \ref{fig:T5-V8}(f)], which is a result of the nonintegrable nature of the final Hamiltonian \cite{srednicki_99,reimann_08}.

\begin{figure}[t]
\centerline {\includegraphics*[width=2.8in]{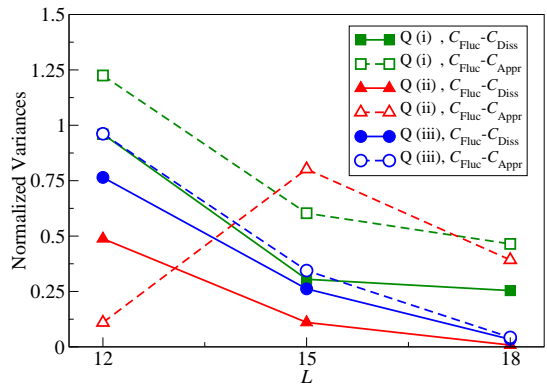}}
\caption{Normalized variance of $C_{\text{Fluc}}(t)-C_{\text{Diss}}(t)$ and 
$C_{\text{Fluc}}(t)-C_{\text{Appr}}(t)$ vs the system size for the three quenches 
explained in the text [identified by {\it Q} (i), {\it Q} (ii), and {\it Q} (iii)], where the normalization 
factor is the average variance of the two functions for which the differences are 
calculated, e.g., Var$(C_{\text{Fluc}}-C_{\text{Diss}}) /\frac{1}{2}[\textrm{Var}
(C_{\text{Fluc}})+\textrm{Var}(C_{\text{Diss}})]$. The observable is $n_{j=L/2}$.
The variances are calculated for 2000 points between $t=0$ and 100.}
\label{fig:var}
\end{figure}

To quantify the differences between the three correlation functions and explore their 
dependence on the system size for each quench type, we calculate the normalized variances
of $C_{\text{Fluc}}(t)-C_{\text{Diss}}(t)$ and $C_{\text{Fluc}}(t)-C_{\text{Appr}}(t)$. 
In Fig.~\ref{fig:var}, we show these quantities for the three quench types vs $L$. 
For quench type (i), the variances exhibit a tendency to saturate to a 
nonzero value as $L$ increases, which indicates that $C_{\text{Fluc}}(t)$ 
and $C_{\text{Diss}}(t)$, as well as $C_{\text{Fluc}}(t)$ and $C_{\text{Appr}}(t)$, 
may remain different in the thermodynamic limit. This is consistent with the 
findings in Refs.~\cite{a_mitra_11,l_foini_11}, where it was
shown that in the thermodynamic limit, conventional fluctuation-dissipation 
relations with a unique temperature do not hold in integrable systems. 
For quench type (ii), we see that the variance of $C_{\text{Fluc}}(t)-C_{\text{Diss}}(t)$ 
decreases with increasing system size and becomes very small already for $L=18$, 
indicating that $C_{\text{Fluc}}(t)$ and $C_{\text{Diss}}(t)$ possibly agree in the
thermodynamic 
limit. The variance of $C_{\text{Fluc}}(t)-C_{\text{Appr}}(t)$, on the other hand,
exhibits a more erratic behavior, and it is not apparent whether it vanishes for larger
system sizes. For quench type (iii), the relative differences between 
$C_{\text{Fluc}}(t)$, $C_{\text{Diss}}(t)$, and $C_{\text{Appr}}(t)$ exhibit a fast 
decline with increasing $L$, indicating that all three likely agree in the 
thermodynamic limit. These results strongly suggest
that the FDT is applicable in the thermodynamic limit for quenches in which the 
final system is nonintegrable, as well as after quenches from nonintegrable to 
integrable systems, even though the ETH does not hold in the latter.

\begin{figure}[t]
\centerline {\includegraphics*[width=3.3in]{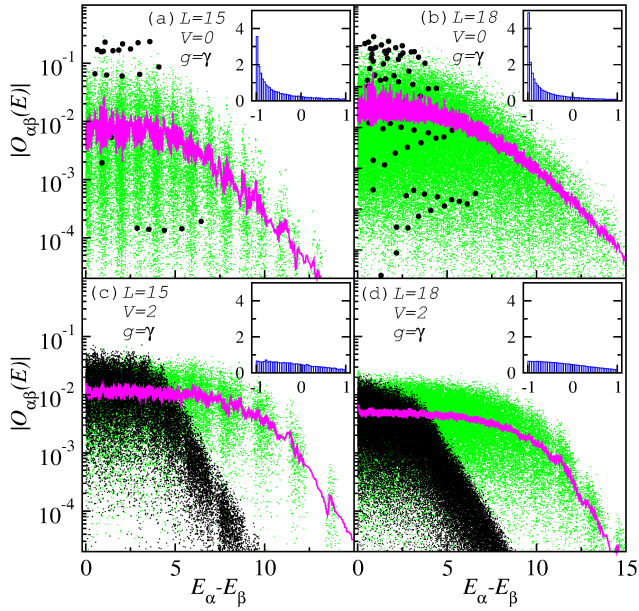}}
\caption{Absolute value of the off-diagonal matrix elements of $\hat{n}_{j=L/2}$ and 
$\hat{n}_{k=0}$ in the eigenenergy basis, in a narrow 
energy window around $E=E_{\textrm{tot}}$ (with a width of 0.1) vs the 
eigenenergy difference $\omega=E_{\alpha}-E_{\beta}$. Results are shown for 
$L=15$ (left panels) and $L=18$ (right panels). (a),(b) and (c),(d) correspond to the final 
Hamiltonian in quenches (ii) and (iii), respectively. The green (light gray) symbols are the matrix 
elements of $\hat{n}_{k=0}$, and the black ones of $\hat{n}_{j=L/2}$. 
In (a) and (b), we have increased the size of the 
symbols for $n_{j=L/2}$ by a factor of 20 relative to those for $n_{k=0}$. 
To increase the resolution of the 
distribution of values in the case of $L=18$, where a very large number of data points 
exists, we plot only 1 out of every 10 points for $n_{k=0}$ in (b) and for both 
observables in (d). Lines are running averages for $n_{k=0}$ with a subset length
of 50 for $L=15$ and 200 for $L=18$. Insets show the histograms of the relative 
differences between the $n_{k=0}$ data and running averages ($f_{\textrm{avg}}$) with subset sizes of 
1000 for $L=15$ and 10 000 for $L=18$. The relative difference is defined as 
$(|O_{\alpha\beta}|-f_{\textrm{avg}})/f_{\textrm{avg}}$.}
\label{fig:off}
\end{figure}

In order to gain an understanding of why FDT fails or applies depending on the 
nature of the final Hamiltonian, we explore to which extent Eq.~\eqref{eq:chaotic} describes
the behavior of the matrix elements of few-body observables in the nonintegrable
case and in which way it breaks down at integrability. In Fig.~\ref{fig:off}, 
we plot the off-diagonal elements of two observables $n_{j=L/2}$ and
the zero-momentum occupation number $n_{k=0}$ 
vs the eigenenergy differences ($\omega$) in a narrow energy window around 
$E=E_{\textrm{tot}}$. Results are shown for matrix elements in the eigenstates of the 
final Hamiltonians in quenches type (ii) and (iii) \cite{note5}. The off-diagonal 
matrix elements of both observables in the eigenstates of the integrable Hamiltonian 
[Figs.~\ref{fig:off}(a) and \ref{fig:off}(b)] exhibit a qualitatively different
behavior from those in the nonintegrable one. In the integrable Hamiltonian, 
they exhibit extremely large fluctuations. 
In addition, a very large fraction of those elements (larger for $n_{j=L/2}$
than for $n_{k=0}$) have vanishing values. This makes any definition 
of a smooth function $f(E,\omega)$ meaningless. Those results contrast the ones
obtained in the nonintegrable case, where the fluctuations of the matrix elements
have a different nature, and we do not find a large fraction of vanishing ones. 
To see that more clearly for $n_{k=0}$ (the better behaved of the two observables), 
in the insets of Fig.~\ref{fig:off}, we show the normalized histograms of the relative 
differences between the matrix elements for $n_{k=0}$ and a ``smooth'' function,
defined as the running average of those elements over a 
large enough group of them (examples of the running averages are presented in 
the main panels). For the integrable system, we find that the histograms are not 
compatible with the uniform distribution postulated in
Eq.~\eqref{eq:chaotic}, as a very sharp peak develops at $-1$ for both system sizes.
That peak becomes sharper with increasing system size, reflecting an 
increasing fraction of vanishing off-diagonal matrix elements in those systems. 
For the nonintegrable Hamiltonian, on the other hand, the histograms are closer to a
uniform distribution.

In summary, studying the dynamics of an experimentally relevant model of trapped hard-core 
bosons with dipolar interactions, we have found indications that the FDT
is applicable to the properties of few-body observables in nonintegrable isolated 
quantum systems out of equilibrium, and that this follows from the ETH. Furthermore, we 
find indications that the FDT may also apply to integrable systems, 
for which the ETH is not valid, provided that the initial state before the quench is an 
equilibrium state (eigenstate) of a nonintegrable system.

We acknowledge support from the National Science Foundation, under Grants No.~OCI-0904597
(E.K. and M.R.) and No. PHY07-57035 (M.S.), and from the European Commission, under ERC-St 
Grant No. 307688 COLDSIM, UdS, and EOARD (G.P.). We are grateful to Aditi Mitra for motivating 
discussions.

\newpage
\ 
\newpage

\onecolumngrid

%\vspace*{0.4cm}

\begin{center}

{\large \bf Supplementary Materials:
\\ Fluctuation-Dissipation Theorem in an Isolated System of Quantum Dipolar Bosons after a Quench}\\

\vspace{0.6cm}

Ehsan Khatami$^{1}$, Guido Pupillo$^2$, Mark Srednicki$^3$, and Marcos Rigol$^4$\\
\ \\
$^1${\it Department of Physics, University of California, Santa Cruz, California 95064, USA}\\
$^2${\it IPCMS (UMR 7504) and ISIS (UMR 7006), Universit\'e de Strasbourg and CNRS, Strasbourg, France}\\
$^3${\it Department of Physics, University of California, Santa Barbara, California 93106, USA}\\
$^4${\it Department of Physics, The Pennsylvania State University, University Park, Pennsylvania 16802, USA}\\

\end{center}

\vspace{0.6cm}

\twocolumngrid

\section{Experimental relevance of our model}

Hamiltonian (1) in the main text
provides a microscopic description for the dynamics of a gas of bosonic ground state molecules such as, e.g.,
LiCs molecules (dipole moment $d_{\rm max} \approx 5.6$ Debye), confined transversely (longitudinally) by a
two-dimensional (one-dimensional) optical lattice with frequency $\omega_\perp$ ($\omega_\parallel$), with
$\omega_\perp \gg \omega_\parallel$. The molecules are polarized in the transverse direction by an external
electric field of strength $F$ and are confined to the lowest band of the 1D lattice, provided
$\omega_\parallel > {\rm max}\{V,J,T\}$. Here, $V= d^2/(4\pi\epsilon_0 a_\parallel^3)$, with
$d \lesssim d_{\rm max}$ the dipole moment induced by $F$, $a_\parallel$ the lattice spacing in 1D and
$\epsilon_0$ the vacuum permittivity. The hard-core condition is obtained by requiring
that molecules are trapped with a low-density $n$, such that the initial system has no doubly occupied
sites~\cite{buchler_micheli_07}. The additional condition $n^{-1/2} \gg (d^2_{\rm max}/\hbar \omega_\parallel)^{1/3}
\simeq 360$nm ensures collisional stability~\cite{buchler_micheli_07}.
The model in Eq. (1) of the main text can also describe the dynamics of a gas of strongly
magnetic atoms such as Dy (dipole moment $d=10\mu_B$, with $\mu_B$ Bohr's magneton) or Er ($d=7\mu_B$).
In this case, the hard-core constraint is achieved by means of magnetic tuning of the short-range scattering
length using Feshbach resonances~\cite{berninger_zenesini_12}, while $J/V$ decreases exponentially 
with increasing the depth of the 1D lattice~\cite{bloch_dalibard_review_08}.

\section{Thermalization}

Despite the presence of interactions that have a power-law decay with distance, we find that
the behavior of eigenstate expectation values of few-body observables, as well as thermalization properties
of the systems described by Hamiltonian (1) of the main text, are qualitatively similar to those 
already seen in models with short-range (nearest and next-nearest-neighbor) interactions 
\cite{rigol_09a,neuenhahn_marquardt_12}. 

In order to show that this is indeed the case, here we study the difference
between the results of the diagonal ensemble for the few-body observables studied in main text, 
namely, $n_{j}$ and $n_{k}$, as well as for the density-density structure factor, $N_{k}$
(not studied in the main text), 
and the results of the microcanonical ensemble. The momentum distribution function and the 
density-density structure factor are defined as
\begin{equation}
\hat{n}_k=\frac{1}{L}\sum_{l,m} e^{ik(l-m)} 
\hat{b}^{\dagger}_l\hat{b}^{}_m,\ \ \hat{N}_k=\frac{1}{L}\sum_{l,m} e^{ik(l-m)} 
\hat{n}^{}_l\hat{n}^{}_m.
\label{eq:momdist}
\end{equation}
They are the Fourier transforms of the one-particle and density-density 
correlation matrices, respectively. Since we work at fixed number of particles,
$\left<N_{k=0}\right>=p^2/L$, so we set it to zero without any loss of generality.
These observables can be studied in ultracold gases experiments. 

\begin{figure}[!b]
\centerline {\includegraphics*[width=3.3in]{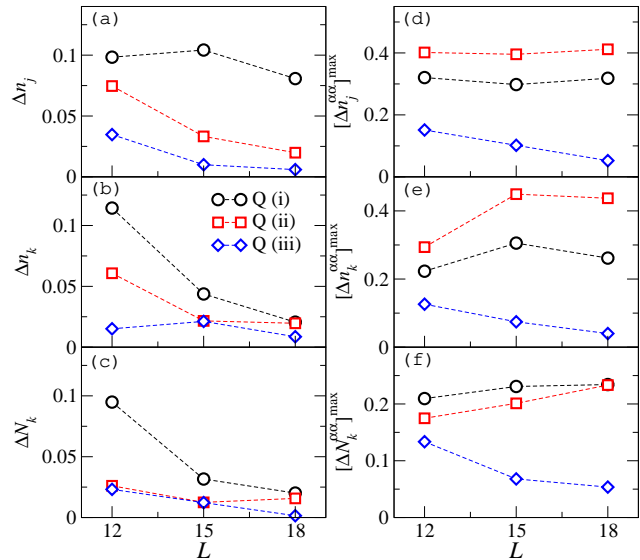}}
\vspace{-0.15cm}
\caption{(a)-(c) Normalized differences between the diagonal and microcanonical 
ensemble averages for the three observables, $n_{j}$, $n_{k}$, and $N_{k}$
[see Eq.~\eqref{eq:therm}] vs $L$. Different plots in each panel correspond to the three 
quench types [identified by Q (i), Q (ii), and Q (iii)]. (d)-(f) Maximum of the normalized
differences between the microcanonical average and each of the eigenstate expectation 
values in the microcanonical energy window [see Eq.~\eqref{eq:ext}] vs $L$.}
\label{fig:delta}
\end{figure}

We define the microcanonical average for an observable $\hat{O}$ as $O_\textrm{micro} 
= \frac{1}{{\cal N}_{\Delta E}}\sum_\alpha O_{\alpha\alpha}$. Here, ${\cal N}_{\Delta E}$ is the 
number of states in the microcanonical window, which is centered around $E_{\textrm{tot}}$ and has a
width of $\Delta E$~\cite{note8}.
We average the results over several close values of $\Delta E$ for each system size
to ensure that they are robust against small changes in $\Delta E$. 
The values for $\Delta E$ are in
$[0.2-0.25]$ for $L=12$, and $[0.1-0.15]$ for $L=15$ and 18. 

We compute the normalized difference between the predictions of the 
diagonal and the microcanonical ensembles,
\be 
\label{eq:therm}
\Delta O =\frac{\sum_\ell |O_\textrm{diag}(\ell) - O_\textrm{micro}(\ell)|}{\sum_l O_\textrm{diag}(\ell)},
\ee
where $\ell$ stands for either the site index or the momentum index depending of the observable.

\begin{figure}[!t]
\centerline {\includegraphics*[width=3.3in]{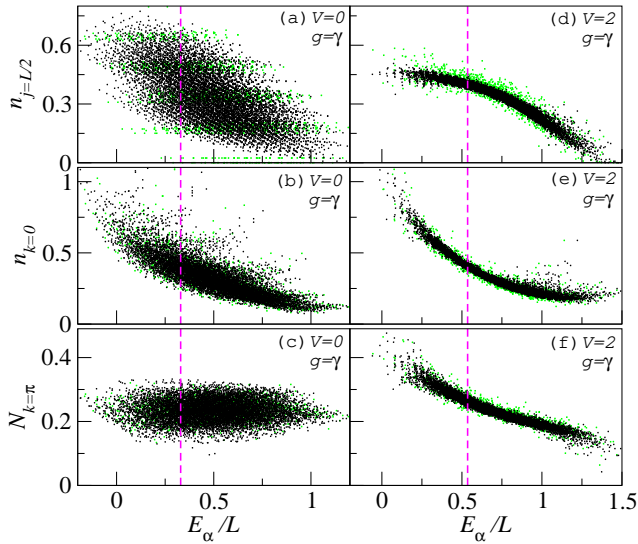}}
\vspace{-0.25cm}
\caption{Eigenstate expectation values values of $\hat{n}_{j=L/2}$, $\hat{n}_{k=0}$, and 
$\hat{N}_{k=\pi}$ over the entire spectrum vs the scaled eigenenergies for the noninteracting 
(a)-(c), and interacting system (d)-(f). We show results for $L=15$ as the green (light colored) 
points, and for $L=18$ as the black (dark) ones in each panel. The vertical lines show the 
location of $E_{\textrm{tot}}/L$ for $L=18$, which is also roughly equal to that of $L=15$.}
\label{fig:eth}
\end{figure}

In Fig.~\ref{fig:delta}(a)--(c), we show results for $\Delta O$ vs $L$ for the three types of quenches
described in the main text. We find that for quench type (i) (between two integrable systems), 
$\Delta O$ is generally larger in comparison to $\Delta O$ for the other two quenches, and that this 
feature remains apparent as the system size increases to $L=18$. Previous studies, involving the same 
integrable model without \cite{rigol_dunjko_07,rigol_muramatsu_06,cassidy_clark_11} and with 
\cite{rigol_muramatsu_06} the trapping potential and much larger system sizes, have found 
strong indications that $\Delta O$ (where $O$ was either the density or the momentum distribution 
function) remains finite in the thermodynamic limit so that the system does not 
thermalize. On the other hand, trends for quench type (iii) suggest that the 
differences vanish by increasing the system size, i.e., that the nonintegrable 
dipolar system thermalizes in the thermodynamic limit. The trend is less clear for 
quench type (ii) as we see that $\Delta n_{j}$ monotonically decreases by increasing 
$L$, whereas $\Delta n_{k}$ and $\Delta N_{k}$ appear to saturate by increasing
$L$ from $L=15$ to $L=18$. Clearly, larger system sizes are required to confirm whether 
thermalization takes place for such an integrable system after a quench from a nonintegrable
one as argued in Ref.~\cite{rigol_srednicki_12}.

As mentioned in the main text, for a generic (nonintegrable) system, thermalization 
can be understood through the ETH. Here, we examine the validity of the ETH for each 
case considered by calculating the normalized differences between the observable in 
each eigenstate and the microcanonical average,
\be 
\label{eq:ext}
\Delta O_{\alpha\alpha} = \frac{\sum_l |O_{\alpha\alpha}(l) - O_\textrm{micro}(l)|}
{\sum_l O_\textrm{micro}(l)},
\ee
and taking the maximal difference within the microcanonical window, 
$[\Delta O^{\alpha\alpha}]^\textrm{max}=\text{Max}[\Delta O_{\alpha\alpha}]_{\Delta E}$. This 
is a measure of how widely the eigenstate expectation values are spread in the microcanonical window.
In Fig.~\ref{fig:delta}(d)-\ref{fig:delta}(f), we show this quantity for our three observables vs $L$.
They are seen to consistently decrease with increasing system size for the final Hamiltonian
in quench type (iii) while they are seem to saturate to relatively large values for the final 
Hamiltonians in quenches type (i) and type (ii). This is an indication that
eigenstate thermalization occurs in the former case while it fails in the latter ones.

The validity, or failure, of ETH can be perhaps more easily seen by plotting the eigenstate 
expectation value of observables vs the eigenenergies, as shown in
Fig.~\ref{fig:eth}. For the integrable system with $V=0$ [Fig.~\ref{fig:eth}(a)-\ref{fig:eth}(c)], 
eigenstate expectation values exhibit large fluctuations inside the microcanonical energy window 
for both $L=15$ and 18, and the width of the region where the values are scattered around 
$E_{\textrm{tot}}$ does not decrease with increasing the system size, i.e., eigenstate thermalization
does not occur. This is different from what happens in the nonintegrable case 
[Fig.~\ref{fig:eth}(d)-\ref{fig:eth}(f)], where each of the eigenstate expectation values 
of an observable inside a narrow energy window around $E_{\textrm{tot}}$ approaches the 
microcanonical average as the system size is increased. This is apparent as
the width of the region where the values reside around $E_{\textrm{tot}}$ 
decreases with increasing system size, and presumably vanishes in the thermodynamic limit. 

\section{energy scales}

The short-time evolution of the correlation functions is set by the width of $f(E,\omega)$ 
as a function of $\omega$, which we denote as $W$. Note that, as discussed in the main text, 
$f$ is not well defined for integrable systems and, even for the nonintegrable case 
for which Eq. (3) of the main text is seen to better describe the data as the system size 
increases, one cannot disentangle $f$ from the random function $R_{\alpha\beta}$ with 
merely the knowledge of the off-diagonal values. Therefore, we estimate $W$ for another 
closely related function, $f_{\textrm{cg}}(\omega)$, which is obtained by coarse-graining 
the off-diagonal values. We then calculate the width using
\be
W=\frac{2\int_0^{\infty}|f_{\textrm{cg}}(\omega)|^2d\omega}{|f_{\textrm{cg}}(0)|^2}.
\label{eq:w}
\ee
Examples of $f_{\textrm{cg}}(\omega)$ are depicted as lines in the right panels of 
Fig.~\ref{fig:fdt12}. We choose different bin sizes for different systems. They are,
$50\times\delta$ for $L=12$, $300\times\delta$ for $L=15$, and $700\times\delta$ for $L=18$,
where $\delta$ is the average level spacing.

\begin{figure}[!t]
\centerline {\includegraphics*[width=3.3in]{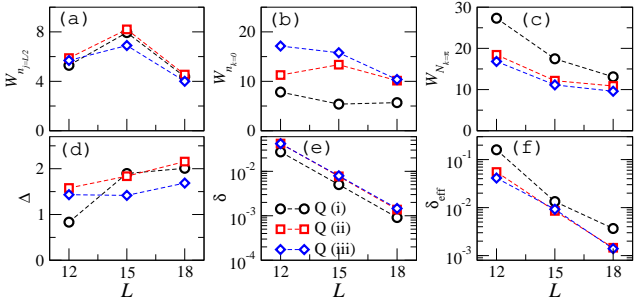}}
\vspace{-0.25cm}
\caption{Various energy scales in the model vs $L$. Top panels show the estimates
for the width of the off-diagonal function, $f(E,\omega)$ in the $\omega$ space.
(d) shows the uncertainty in the total energy, and (e)-(f) show the average level 
spacings (see text).}
\label{fig:deltas}
\end{figure}

In Fig.~\ref{fig:deltas}(a)-\ref{fig:deltas}(c), we plot $W$ versus $L$ in the three 
quenches and for the three observables considered. For $n_{j=L/2}$ and $n_{k=0}$, 
$W$ varies non-monotonically with $L$ in most cases, while for $N_{k=\pi}$, it is seen 
to monotonically decrease with increasing $L$ and possibly saturate to a finite value
for larger system sizes. Larger system sizes are needed to understand the behavior of 
$W$ in macroscopic systems. Regardless, the values obtained here
can be used in each case to estimate the time scale for the initial decrease of the 
fluctuation or dissipation correlation functions. 

In Fig.~\ref{fig:deltas}(d), we also show the quantum uncertainty of the energy, 
$\Delta$. We find that they are $\mathcal{O}(10^{0})$ for the systems studied here, 
and increase slowly with system size, as expected from the analysis in 
Ref.~\cite{rigol_dunjko_08}.

The time scale for recurrences in the correlation functions is set by the average 
level spacing, $\delta$ (estimated by $e^{S(E)}E$~\cite{srednicki_99}), which, as expected, 
is found to decrease 
exponentially with increasing the system size from $\mathcal{O}(10^{-1})$ for $L=12$ to 
$\mathcal{O}(10^{-3})$ for $L=18$ [see Fig.~\ref{fig:deltas}(e)]. Therefore, such a 
time scale for typical system sizes explored in experiments would be much too large 
to have any relevance. This is also true if one calculates $\delta_{\textrm{eff}}$
($=\frac{\Delta}{\sum_{\alpha}|c_{\alpha}|^4}$)~\cite{srednicki_99}, shown in 
Fig.~\ref{fig:deltas}(f), which represents the effective level spacing between the 
eigenstates participating in the diagonal ensemble.

\section{Other observables and/or system size}

\begin{figure}[!h]
\centerline {\includegraphics*[width=3.3in]{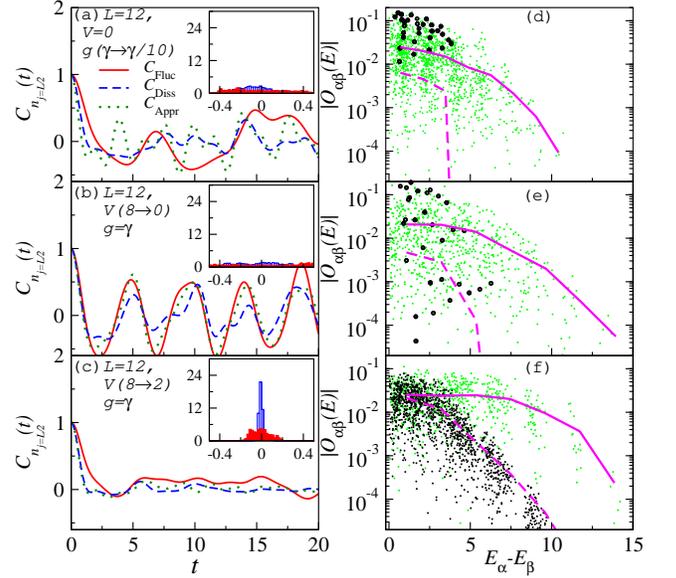}}
\vspace{-0.25cm}
\caption{Left panels (Right panels): the same as Fig. 1 (Fig. 3) of the main text, 
except that $L=12$, and that the lines in the right panels are coarse-grained 
functions of the off-diagonal values used to calculate the width of data in $\omega$
using Eq.~\eqref{eq:w}. In the right panels, solid (dashed) lines correspond to $n_{k=0}$ ($n_{j=L/2}$).
The bin size for coarse-graining is $50\times\delta$ for this system size.}
\label{fig:fdt12}
\end{figure}

In the main text, we show the fluctuation and dissipation correlation functions
of $n_{j=L/2}$ for the two largest system sizes, $L=15$ and 18. For completeness, 
in Fig.~\ref{fig:fdt12}, we show the same quantities, as well as the corresponding off-diagonal 
elements of the two observables shown in Fig. 3 of the main text, for the smallest system
we have studied, $L=12$. Because of the smaller size of the Hilbert space in 
comparison to the other clusters, some of the trends seen in Fig. 1 of the main 
text are not so clear in the case of $L=12$. However, the suppression in the fluctuations 
of the correlation functions for the nonintegrable case [Fig.~\ref{fig:fdt12}(c)] is 
apparent in the corresponding histogram. Other features, such as the dramatic change 
in the behavior of the off-diagonal elements of $n_{j=L/2}$ when the integrability-breaking
interaction is introduced, is already seen in Figs.~\ref{fig:fdt12}(d)-\ref{fig:fdt12}(f) 
for this small cluster.

\begin{figure*}[!t]
\centerline {\includegraphics*[width=4in]{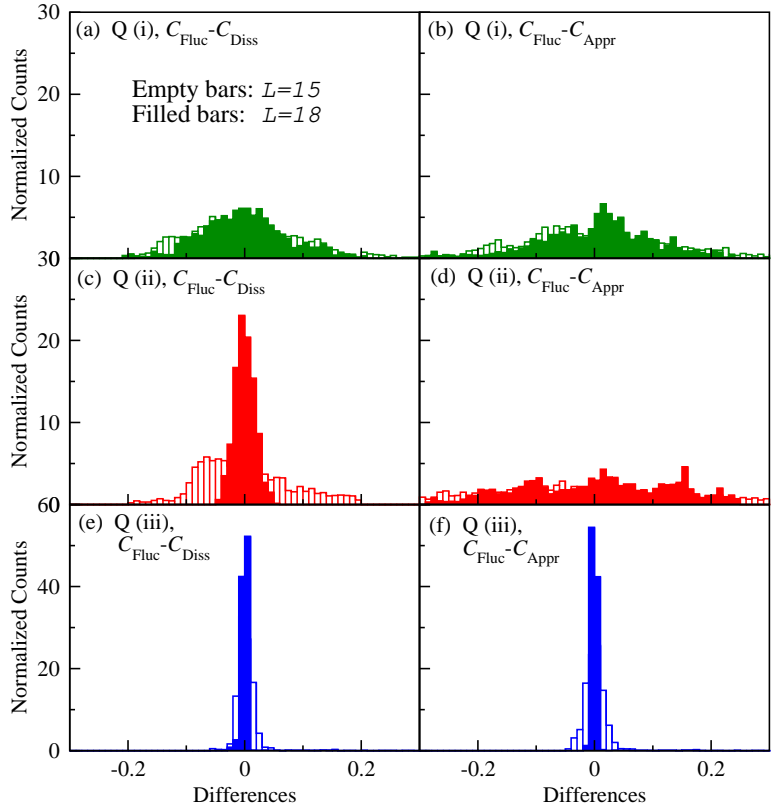}}
\caption{Normalized histograms of $C_{\text{Fluc}}(t)-C_{\text{Diss}}(t)$ (left panels)
and $C_{\text{Fluc}}(t)-C_{\text{Appr}}(t)$ (right panels) for $n_{j=L/2}$, and for the
three quenches (from top to bottom) and the two largest system sizes, calculated for 2000
data points between 
$t=0$ and 100.}
\label{fig:diff}
\end{figure*}

In Fig.~\ref{fig:diff}, we show the histograms of the differences between
$C_\text{Fluc}(t)$ and $C_\text{Diss}(t)$, and between $C_\text{Fluc}(t)$ and
$C_\text{Appr}(t)$ for the three quench types studied in the main text and for the two
largest system sizes accessible to us. The results in this figure complement those of
the normalized variances presented in Fig. 2 of the main text. Figures ~\ref{fig:diff}(a)
and ~\ref{fig:diff}(b) show that, in quenches type (i), there are no signatures of a 
reduction with increasing system size of the large differences seen between the different
correlation functions at each given time (Fig. 1 of the main text). A similar conclusion
stands for the behavior of $C_\text{Fluc}(t)-C_\text{Appr}(t)$ in quenches type (ii)
[Fig.~\ref{fig:diff}(d)]. On the other hand, the histograms of
$C_\text{Fluc}(t)-C_\text{Diss}(t)$ in quenches type (ii) [Fig.~\ref{fig:diff}(c)], and of
$C_\text{Fluc}(t)-C_\text{Diss}(t)$ as well as $C_\text{Fluc}(t)-C_\text{Appr}(t)$ in
quenches type (iii) [Fig.~\ref{fig:diff}(e) and ~\ref{fig:diff}(f), respectively] make
apparent than not only does the variances decrease with increasing system size as shown
in Fig. 1 of the main text, but also the maximal differences between the correlation
functions at each given time decrease with increasing system size.

\begin{figure*}[!t]
\centerline {\includegraphics*[width=5.5in]{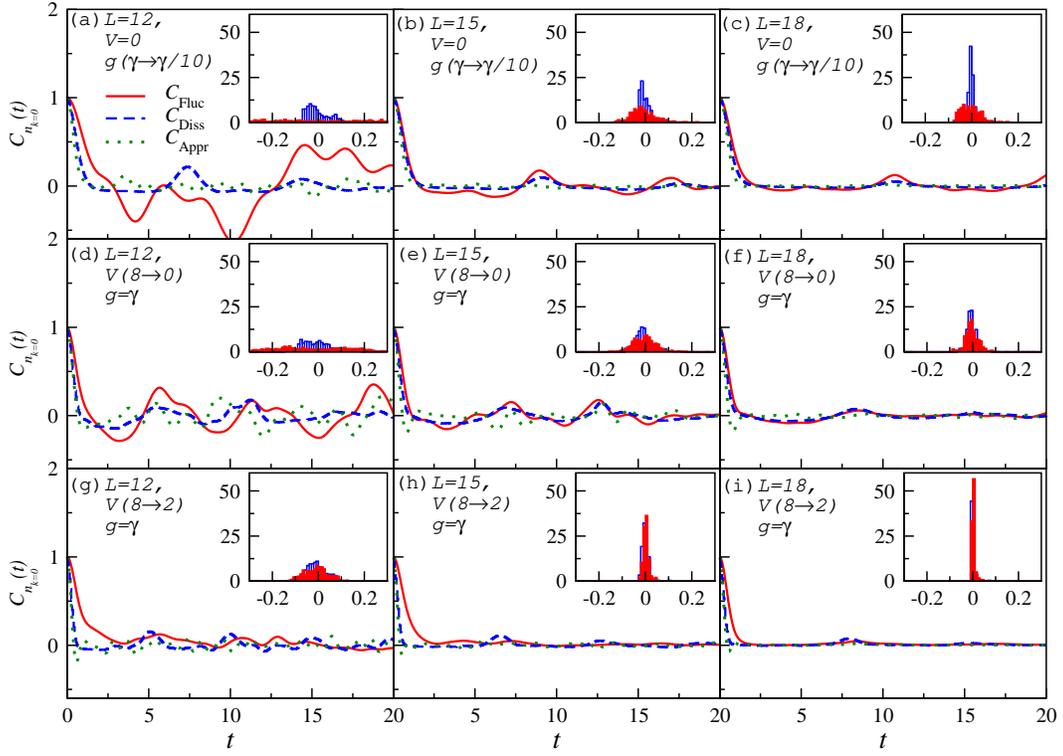}}
\caption{The same as Fig. 1 of the main text, except that here, the observable is
the zero-momentum occupation number, $n_{k=0}$. We have also included the results
for $L=12$ in the left panels.}
\label{fig:fdt_nk}
\end{figure*}

\begin{figure*}[!b]
\centerline {\includegraphics*[width=5.5in]{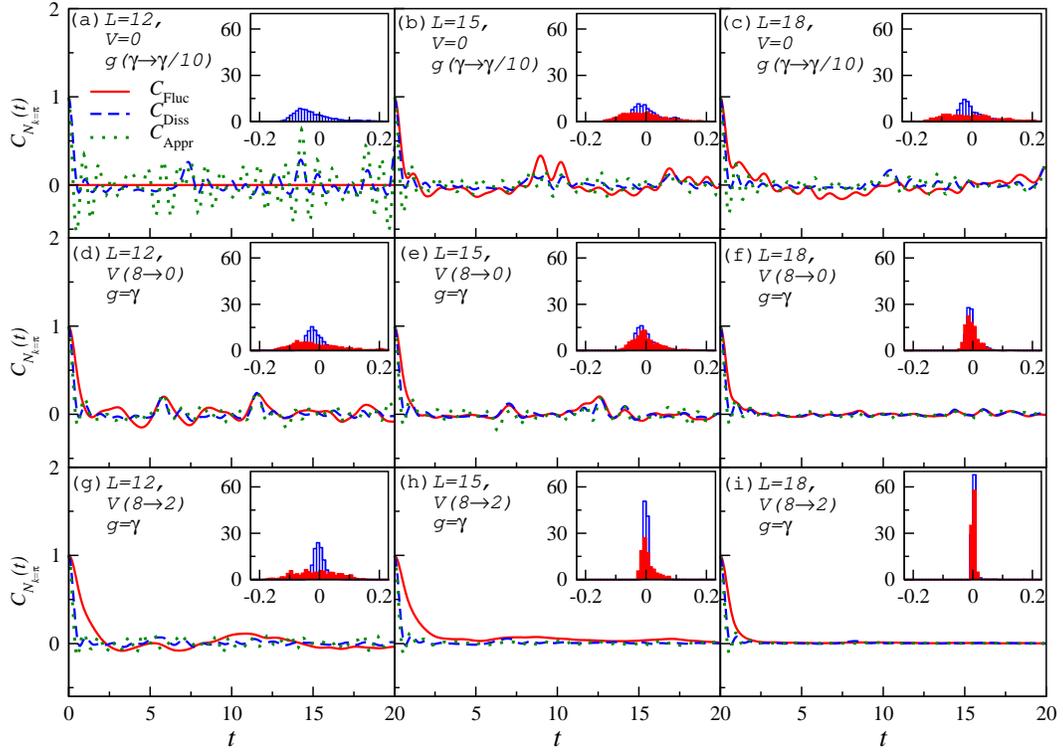}}
\caption{The same as Fig. 1 of the main text, except that here, the observable is
the $k=\pi$ density-density structure factor, $N_{k=\pi}$. We have also included 
the results for $L=12$ in the left panels. For quench type (i), we find that the value 
of $N_{k=\pi}$ for $L=12$ does not exhibit any dynamics. This is reflected in the 
corresponding fluctuation correlation function in (a), which is zero at all times. For 
this reason, we have set $C_{\text{Fluc}}(t)=0$ in that case and are not showing
its histogram. This is unique to $N_{k=\pi}$ for $L=12$ and to the specific
parameters chosen for quench type (i).}
\label{fig:fdt_Nk}
\end{figure*}

In Figs.~\ref{fig:fdt_nk} and \ref{fig:fdt_Nk}, we show results for the fluctuation and 
dissipation correlation functions of $n_{k=0}$ and $N_{k=\pi}$ for the three system sizes 
and quench types considered in the main text. The contrast between the results for different
quenches can be seen to be similar to the one in Fig. 1 of the main text for $n_{j=L/2}$.
The main difference between the results for $n_{k=0}$ and $N_{k=\pi}$ when compared to those
 for $n_{j=L/2}$ is that the former two exhibit smaller fluctuations with increasing 
system size than the latter one. This is to be expected as the presence of the harmonic
trap, which breaks translational symmetry, produces a larger number of nonzero off-diagonal
matrix elements of $n_{k=0}$ and $N_{k=\pi}$ in the integrable regime than of $n_{j=L/2}$
(see Fig. 3 of the main text). Still, those fluctuations [see the histograms in the insets of
Figs.~\ref{fig:fdt_nk}(a)--\ref{fig:fdt_nk}(f) and Figs.~\ref{fig:fdt_Nk}(a)--\ref{fig:fdt_Nk}(f)] 
can be seen to be much stronger than in the nonintegrable case 
[see the histograms in the insets in 
Figs.~\ref{fig:fdt_nk}(g)--\ref{fig:fdt_nk}(i) and Figs.~\ref{fig:fdt_Nk}(g)--\ref{fig:fdt_Nk}(i)].

\begin{figure*}
\centerline {\includegraphics*[width=5.3in]{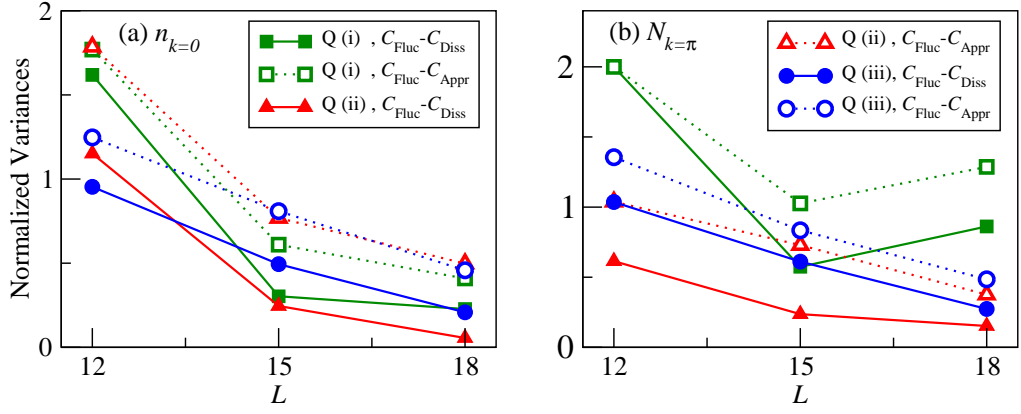}}
\caption{Same as Fig. 2 of the main text, but for when the observables are $n_{k=0}$ (a),
and $N_{k=\pi}$ (b).}
\label{fig:var2}
\end{figure*}

The results for the scaling of the variances between $C_{\text{Fluc}}(t)-C_{\text{Diss}}(t)$ 
and $C_{\text{Fluc}}(t)-C_{\text{Appr}}(t)$ for $n_{k=0}$ and $N_{k=\pi}$ are presented in 
Fig.~\ref{fig:var2}. They are also consistent with the conclusions extracted from the scaling 
of those differences for $n_{j=L/2}$.

\begin{figure*}
\centerline {\includegraphics*[width=5.3in]{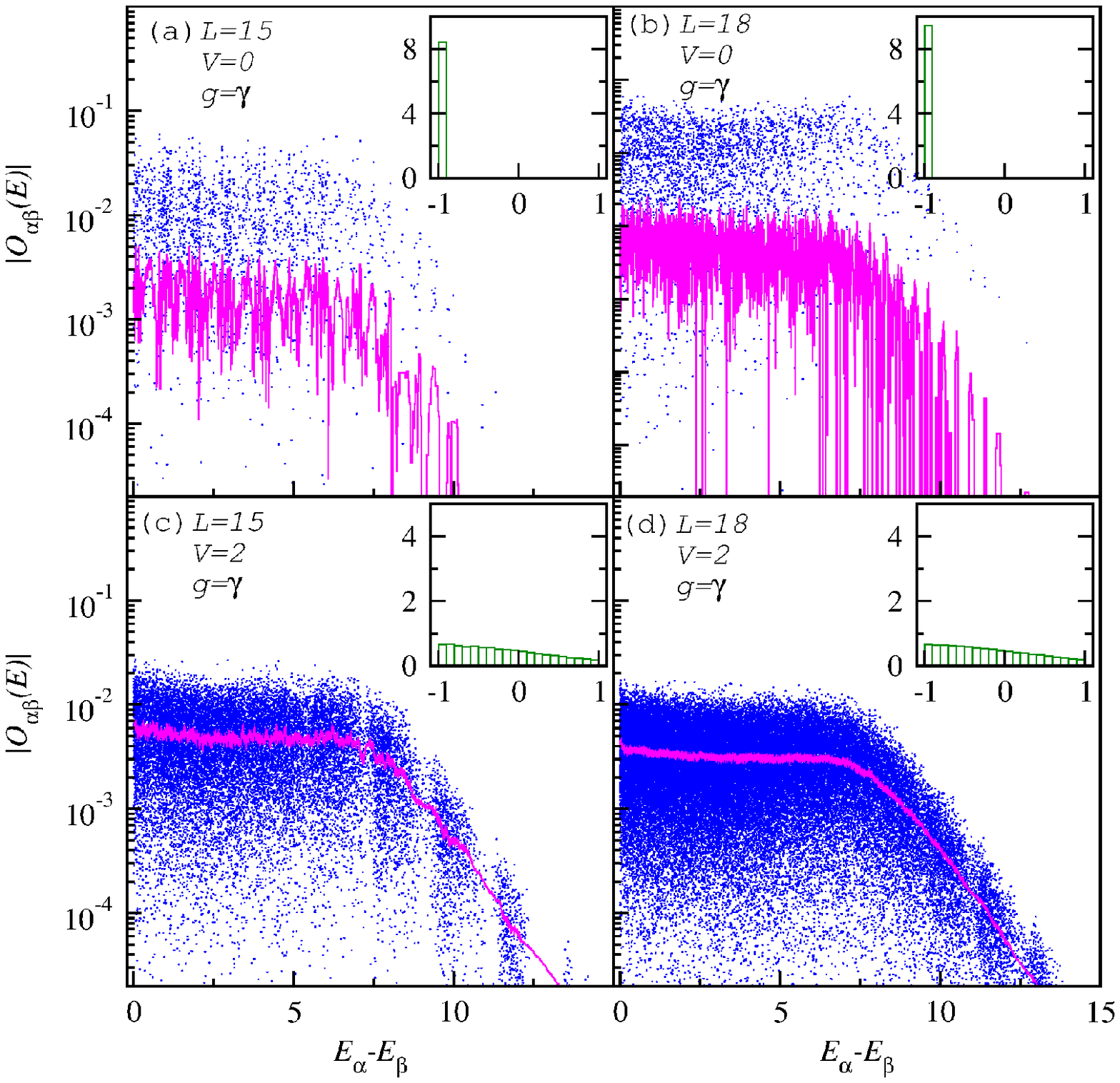}}
\caption{Same as Fig. 3 of the main text, but for when the observable is $N_{k=\pi}$. 
Here, we show all the data points in all of the cases. Lines are running averages 
with a subset length of 100 for $L=15$ and 1000 for $L=18$.}
\label{fig:off2}
\end{figure*}

Finally, in Fig.~\ref{fig:off2}, we show results for the off-diagonal matrix elements of 
$N_{k=\pi}$ within $\Delta E$ of $E_{\textrm{tot}}$. Those results are the equivalent of the
ones presented in Fig. 3 of the main text for $n_{j=L/2}$ and $n_{k=0}$. That figure shows that 
the conclusions drawn for the latter two observables in the main text are also applicable to 
$N_{k=\pi}$.

\begin{thebibliography}{72}
\bibitem[{not({\natexlab{e}})}]{note7}
\bibinfo{note}{Note that $t'$ is not the time after the quench. The question of interest
here is whether fluctuation-dissipation relations hold after the observable of
interest has equilibrated following a sudden quench. We note that most of the
time after the quench, the observable will be close to its infinite-time
average (or equilibrated) value~\cite{reimann_08}. What we study is how observables relax
to the infinite-time averages following a random fluctuation that can occur at
a random time after equilibration.}
\bibitem[{not({\natexlab{b}})}]{note1}
\bibinfo{note}{By ``overwhelmingly'' we mean that the conditional probability
  distribution $P(O_{t'+t}|O_{t'})$ is a Gaussian centered at $C(t)O_{t'}$
  whose width is proportional to $\delta/\Delta$, where $\Delta$ is the
  uncertainty in energy and $\delta$ is the level spacing \cite{srednicki_99}.
  It is expected that in quenches in large systems, $\Delta\gg\delta$.}
\bibitem[{not({\natexlab{c}})}]{note2}
\bibinfo{note}{We have subtracted a constant term containing the infinite-time
  average of $O_t$ from this correlation function before normalization such
  that the infinite-time average of $C_{\text{Fluc}}(t)$ is zero.}
\bibitem[{not({\natexlab{d}})}]{note4}
\bibinfo{note}{We compute $f(E,\omega)$ by considering those off-diagonal matrix 
  elements of observables whose $E$ falls in a narrow window $\Delta E$
  around the total energy $E_{\textrm{tot}}$, and by sorting them by the value of
  $\omega$. $\Delta E$ is chosen such that the results are robust against small
  changes of $\Delta E$ (0.1 for $L=18$ and 15, and 0.2 for $L=12$).}
\bibitem{note6}
We have studied the thermalization properties of the system in more detail~\cite{note3}
and found that they are qualitatively similar to those in models with short-range interactions
\cite{rigol_09a}.
\bibitem[{not({\natexlab{e}})}]{note5}
\bibinfo{note}{The results for the final Hamiltonian in quench type (i) are
  qualitatively similar to those of the final Hamiltonian in quench type (ii),
  and are not reported here.}
\end{thebibliography}
\end{document}